\documentclass[aps,superscriptaddress,showpacs,floatfix,nobibnotes]{revtex4}

\usepackage{subfigure}
\usepackage{epsfig}
\usepackage{epstopdf}
\usepackage{color}
\usepackage{graphicx}
\usepackage{longtable}
\usepackage{amsmath,bbold}
\usepackage{float}

\ifx\pdfoutput\undefined
\usepackage[dvips]{graphicx}
\usepackage[dvipdfm,
  pdfstartview=FitH,
     CJKbookmarks=true,
          bookmarksnumbered=true,
          bookmarksopen=true,
         colorlinks=true,
          pdfborder=002,
          citecolor=blue%
           ]{hyperref}
\AtBeginDvi{} % GBK -> Unicode
\else
\usepackage[
          pdfstartview=FitH,
          CJKbookmarks=true,
          bookmarksnumbered=true,
           bookmarksopen=ture,
           colorlinks=true,
           citecolor=blue%         ¡¡
           ]{hyperref}
\begin{document}

\title{\Large Exact solution of the two-axis two-spin  Hamiltonian\\
\vskip .3cm
{\small Feng Pan$^{1,3}$, Yao-Zhong Zhang$^{2}$\footnote{The corresponding author.\\
E-mail address: yzz@maths.uq.edu.au},
Xiaohan Qi$^{1}$,
Yue Liang$^{1}$,
Yuqing Zhang$^{1}$,
and Jerry P. Draayer$^{3}$} }

\address{Department of Physics, Liaoning Normal University, Dalian 116029, P. R. China\\
$^{2}$School of Mathematics and Physics, The University of Queensland, Brisbane, Qld 4072, Australia\\
$^{3}$Department of Physics and Astronomy, Louisiana State University, Baton Rouge, Louisiana 70803-4001, USA
}

\date{\today}

\begin{abstract}
\noindent Bethe ansatz solution of the two-axis two-spin Hamiltonian  is derived based on the Jordan-Schwinger  boson realization
of the SU(2) algebra. It is shown that the solution of the Bethe ansatz equations can be obtained as zeros of the related extended
Heine-Stieltjes polynomials. Symmetry properties of excited levels of the system and zeros of the related extended
Heine-Stieltjes polynomials are discussed. As an example of an application of the theory,
the two equal spin case is studied in detail, which demonstrates that the levels in each band are symmetric with respect to
the zero energy plane perpendicular to the level diagram and that excited states are always well entangled.

\end{abstract}


\maketitle

\section{INTRODUCTION}

Quantum squeezing of both Bose
and Fermi many-body systems~\cite{1,11,12,13,14,15,17,18,31,32,33,34}
is effective and useful in quantum metrology and its applications in quantum informatics~\cite{222,333}.
Two previous models for dynamical generation of spin-squeezed states are
the one-axis twisting and two-axis countertwisting Hamiltonians, respectively \cite{1}, of which
the latter model gives rise to maximal squeezing with a squeezing angle independent of system size or evolution time. Very recently, two-axis one-spin (2A1S) countertwisting
Hamiltonian \cite{1} has been generalized to the
two-axis two-spin (2A2S) case~\cite{2222}. As shown in \cite{2222}, the 2A2S
Hamiltonian produces the spin EPR states, the analog of the two-mode squeezed
state for spins, which are able to violate the Bell-CHSH inequality
when the quantum numbers of the two spins are finite.

\vskip .3cm
It wss shown in our previous works \cite{pan-zhang1, pan-zhang2}
that the 2A1S Hamiltonian is exactly solvable and its solution can be obtained by using
the Bethe ansatz method. As noted in \cite{pan-zhang2},
the 2A1S Hamiltonian is equivalent to a special case of the Lipkin-Meshkov-Glick (LMG) model~\cite{vi1,vi2} after an Euler rotation.
Similar to other many-spin systems~\cite{be,gau,ric1,ric2},
the LMG model can be solved analytically by using the algebraic Bethe ansatz
\cite{pan, mori}. The same problem can also be solved by using the Dyson boson realization of the SU(2) algebra~\cite{vi1,vi2,zhang1,zhang2},
of which the solution may be obtained from the Riccati differential equations~\cite{vi1,vi2}.
Recently, the Bethe ansatz method has also been applied
to generate exact solution of mean-field plus orbit-dependent non-separable pairing
model with two non-degenerate $j$-orbits \cite{pan2019}
and that of the dimer Bose-Hubbard model with
multi-body interactions \cite{pan2020}. In these two models \cite{pan2019, pan2020},
there is an additional $S_{2}$ symmetry
with respect to the two species of bosons or quasi-spins, which
is helpful in constructing operators involved in the
corresponding Bethe ansatz states and the related
operator algebra calculations.

\vskip .3cm
The purpose of this work is to construct exact and complete solution of the 2A2S Hamiltonian. In Sec. II, the 2A2S Hamiltonian is written
in terms of boson operators  after the Jordan-Schwinger
boson realization of the two related SU(2) algebras, which can then be expressed
in terms of  generators of two copies of SU(1,1) algebra.
Since there are only two sets of SU(1,1) generators
involved in the Hamiltonian, the technique used in \cite{pan2019,pan2020}
is helpful in constructing the Bethe ansatz states and the related operator algebra
calculations. The derivation of the related extended Heine-Stieltjes polynomials
is presented, whose zeros are related to the exact solution of the 2A2S Hamiltonian.
Symmetry properties of excited levels of the system
and zeros of the related extended Heine-Stieltjes polynomials
are also discussed. Some numerical examples of the two equal spin case
are presented in Sec. III, which demonstrate the main
features of the solution. A brief summary is provided in Sec. IV.

\section{The exact solution of the two-axis two-spin Hamiltonian}

The two-axis two-spin (2A2S) Hamiltonian is given by~\cite{2222}
 \begin{equation}\label{1}
 {H}_{\rm 2A2S}={\chi}({S}^{+}_{1}S^{+}_{2}+S^{-}_{1}S^{-}_{2}),
 \end{equation}
where $\chi$ is  a  constant and
$\{S^{\pm}_{i},S^{0}_{i}\}$ ($i=1,2$) are generators of two copies of the SU(2)
algebra satisfying the following commutation relations:
\begin{eqnarray}\label{su2}
[S^{0}_{i},~ S^{\pm}_{j}]=\pm\delta_{ij}S^{\pm}_{i},~~
[S^{+}_{i},~ S^{-}_{j}]=\delta_{ij}2S^{0}_{i}.
 \end{eqnarray}
Though the Hamiltonian (\ref{1}) is not commutative
with $S^{\nu}_{i}$ ($\nu=+,\,-,\,0;\,i=1,\,2$), it is commutative
with the two SU(2) Casimir invariants
$C^{(i)}_{2}({\rm SU}(2))=(S^{+}_{i}S^{-}_{i}+S^{-}_{i}S^{+}_{i})/2+(S^{0}_{i})^{2}$
for $i=1,\,2$. Thus, the two spins are good quantum numbers of the system, while the total spin and its projection are not.
\vskip .3cm
The generators of the two copies of the SU(2)
can be represented using the Jordan-Schwinger boson realization
\begin{eqnarray}\label{2s}\nonumber
S^{+}_{1}=a^{\dag}_{1}a_{2},~~ S^{-}_{1}=a^{\dag}_{2}a_{1},~S^{0}_{1}={1\over{2}}(a^{\dag}_{1}a_{1}-a^{\dag}_{2}a_{2}),\\
S^{+}_{2}=b^{\dag}_{1}b_{2},~~S^{-}_{2}=b^{\dag}_{2}b_{1},~~S^{0}_{2}=
{1\over{2}}(b^{\dag}_{1}b_{1}-b^{\dag}_{2}b_{2}),~
 \end{eqnarray}
where $a_{i}$ and $b_{i}$ ($a^{\dag}_{i}$ and $b^{\dag}_{i}$)
are the boson annihilation (creation) operators satisfying
\begin{eqnarray}
[a_{i},\,a^{\dag}_{j}]= [b_{i},\,b^{\dag}_{j}]=\delta_{ij},~~
[a_{i},\,a_{j}]=[b_{i},\,b_{j}]=[a_{i},\,b_{j}]=0.
 \end{eqnarray}

By substituting (\ref{2s}) into (\ref{1}), the Hamiltoanian (\ref{1})
can be expressed in terms of the generators of two copies of SU(1,1) algebra,
\begin{eqnarray}\label{3}
 {H}_{\rm 2A2S}={\chi}\left(\Lambda_{1}^{+}\Lambda_{2}^{-}+ \Lambda_{1}^{-}\Lambda_{2}^{+}
 \right)
 \end{eqnarray}
with
\begin{eqnarray}\label{2}\nonumber
 \Lambda^{+}_{1}=a^{\dag}_{1}b^{\dag}_{1},~~ \Lambda^{-}_{1}=a
_{1}b_{1},~\Lambda^{0}_{1}={1\over{2}}(a^{\dag}_{1}a_{1}+b^{\dag}_{1}b_{1}+1),\\
\Lambda^{+}_{2}=a^{\dag}_{2}b^{\dag}_{2},~~\Lambda^{-}_{2}=a_{2}b_{2},~~\Lambda^{0}_{2}=
{1\over{2}}(a^{\dag}_{2}a_{2}+b^{\dag}_{2}b_{2}+1),
 \end{eqnarray}
which satisfy the commutation relations
\begin{equation}\label{su11}
[\Lambda^{0}_{\rho},~\Lambda^{\pm}_{\rho^{\prime}}]=\pm\delta_{\rho\rho^{\prime}}\Lambda^{\pm}_{\rho},~~~
[\Lambda^{+}_{\rho},~ \Lambda^{-}_{\rho^{\prime}}]=-\delta_{\rho\rho^{\prime}} 2\Lambda^{0}_{\rho}.
\end{equation}

As shown in the following,
the SU(1,1) type Bethe ansatz states for the Hamiltonian (\ref{3}) can be written as
\begin{equation}\label{su11ba}
\vert \eta, k; \lambda_{1},\lambda_{2}\rangle=
\prod^{k}_{i=1}\Lambda^{+}(x^{(\eta)}_{i})\vert\lambda _{1},\lambda _{2}\rangle,
\end{equation}
where $\eta$ is an additional label needed,
$\vert\lambda _{1},\lambda _{2}\rangle$ is one of the lowest weight states
of ${\rm SU}_{1}(1,1)\otimes \rm{SU}_{2}(1,1)$ satisfying
\begin{equation}\label{su11c}
\left(\begin{array}{c}
\Lambda^{0}_{i}\\
~\Lambda^{-}_{i}\\
\end{array}\right)
\vert  \lambda_{1},\lambda_{2}\rangle=
\left(\begin{array}{c}
\lambda _{i}\\
0\\
\end{array}\right)\vert\lambda _{1},\lambda _{2}\rangle~~{\rm for}~~i=1,2,
\end{equation}
and
\begin{equation}\label{op}
\Lambda^{+}(x)=\Lambda^{+}_{1}+ x \,\Lambda^{+}_{2}
\end{equation}
with variable $x$ to be determined.
The allowed $\lambda_{1}$ and $\lambda_{2}$ values are provided in Table \ref{t1}.

\begin{table}[H]
\caption{The allowed $\lambda_{1}$ and $\lambda_{2}$ values of
the ${\rm SU}_{1}(1,1)\otimes {\rm SU}_{2}(1,1)$
lowest weight states satisfying (\ref{su11}) and
the corresponding boson and the two-spin intrinsic quantum numbers,
where $\mu_{1}$ and $\mu_{2}$ can be taken as arbitrary positive integers or zero.
\label{t1} }
\begin{center}
\begin{tabular}{|c|c|c|c|c|c|c|c|c|c|}
\hline
$\lambda_{1}$&$\lambda_{2}$ &~$n_{a_{1}}$~ &~$n_{a_{2}}$~&~$n_{b_{1}}$ ~&~$n_{b_{2}}$~
&$S^{\rm in}_{1}$&$M^{\rm in}_{1}$ &$S^{\rm in}_{2}$&$M^{\rm in}_{2}$\\
\hline
$~{\mu_{1}+1\over{2}}$~&$~~{\mu_{2}+1\over{2}}~~$ &$\mu_1$&$\mu_{2}$&$0$&$0$ &~~${1\over{2}}(\mu_{1}+\mu_{2})$~~&~~${1\over{2}}(\mu_{1}-\mu_{2})$~~&0&$0$\\
\hline
& &$0$&$0$&$\mu_1$&$\mu_2$ &$0$&0&~~${1\over{2}}(\mu_{1}+\mu_{2})$~~&~~${1\over{2}}(\mu_{1}-\mu_{2})$~~\\
\hline
& &$\mu_1$&$0$&$0$&$\mu_2$ & ${\mu_{1}\over{2}}$&${\mu_{1}\over{2}}$
&${\mu_{2}\over{2}}$&$-{\mu_{2}\over{2}}$\\
\hline
& &$0$&$\mu_2$&$\mu_1$&~$0$ &${\mu_{2}\over{2}}$&$-{\mu_{2}\over{2}}$
&${\mu_{1}\over{2}}$&${\mu_{1}\over{2}}$\\
\hline
\end{tabular}
\end{center}
\end{table}
\vskip .3cm

The  single- and double-commutators of the Hamiltonian (\ref{3})
with the operator (\ref{op}) can be expressed as
\begin{equation}\label{cr1}
[{H}_{\rm 2A2S}/\chi,\Lambda^{+}(x)]=
2x\Lambda^{+}_{1}\Lambda_{2}^{0}+2\Lambda_{2}^{+}\Lambda_{1}^{0},
\end{equation}
\begin{equation}\label{cr2}
[\,[{H}_{\rm 2A2S}/\chi,\Lambda^{+}(x)],\Lambda^{+}(y)]=
2(x\,y+1)\Lambda^{+}_{1}\Lambda_{2}^{+} ,
\end{equation}
while other higher order commutators of the Hamiltonian (\ref{3})
with the operator (\ref{op}) vanish.
Once these commutators are obtained,  the corresponding
polynomials $G_{1}(x)$ and $G_{2}(x,y)$ in $\Lambda^{+}_{1}$ and $\Lambda^{+}_{2}$ on
the lowest weight states
of ${\rm SU}_{1}(1,1)\otimes {\rm SU}_{2}(1,1)$, defined as  \cite{pan2020}
\begin{eqnarray}\label{g}
G_{1}(x)\vert\lambda_{1},\lambda_{2}\rangle &=&[{H}_{\rm 2A2S}/\chi,\Lambda^{+}(x)]\vert\lambda_{1},\lambda_{2}\rangle,\nonumber\\
G_{2}(x,y)\vert\lambda_{1},\lambda_{2}\rangle &=& [\,[{H}_{\rm 2A2S}/\chi,\Lambda^{+}(x)],\Lambda^{+}(y)]\vert\lambda_{1},\lambda_{2}\rangle,
\end{eqnarray}
can be expressed in the form
\begin{eqnarray} 
G_{1}(x)&=&\alpha(x)\,\Lambda^{+}(g)+\beta(x)\,\Lambda^{+}(x),\label{g1}\\
G_{2}(x,y)&=&a(x,y)\, \Lambda^{+}(g)\,\Lambda^{+}(x)+b(x,y)\, \Lambda^{+}(g)\,\Lambda^{+}(y)+c(x,y)\,\Lambda^{+}(x)\,\Lambda^{+}(y).\label{g2}
\end{eqnarray}
Here $g$ is a free parameter whose allowed values will be determined later and
\begin{equation}\label{g1p}
\alpha(x)={2\,(\lambda_{2}\,x^2-\lambda_{1})\over{x-g}},~~~~
\beta(x)={2\,(\lambda_{1}-\lambda_{2}\,x\,g)\over{x-g}},
\end{equation}
\begin{eqnarray}\label{g2p}
a(x,y)&=& {2y(1+x y)\over{(g-y)(y-x)}},~~~~
b(x,y)=a(y,x)= {2x(1+x y)\over{(g-x)(x-y)}},\nonumber\\
c(x,y)&=&c(y,x)=-{2g(1+x y)\over{(g-x)(g-y)}},
\end{eqnarray}
It can be observed that the single- and double-commutators of the Hamiltonian (\ref{3})
with the operator (\ref{op}) are quite similar to the corresponding ones
appearing in the Richardosn-Gaudin type models \cite{pan2019,pan2020}.
Therefore, the SU(1,1) type Bethe ansatz (\ref{su11ba}) works for this case.

\vskip .3cm
Similar to what is shown in \cite{pan2019},
using the commutation relations (\ref{cr1}), (\ref{cr2}),
and the expressions shown in (\ref{g1}) and (\ref{g2}),
we can directly check that
\begin{eqnarray}\label{15}
({H}_{\rm 2A2S}/\chi)\vert \eta,~k;\lambda_{1},\lambda_{2}\rangle &=&\sum^{k}_{i=1}
\alpha(x^{(\eta)}_{i})\,\Lambda^{+}(g)\prod_{\rho\,(\neq i)}^{k}\Lambda^{+}(x^{(\eta)}_{\rho}) \vert \lambda_{1},\lambda_{2}\rangle \nonumber\\
& &+\sum^{k}_{i=1}\beta(x^{(\eta)}_{i})\prod_{\rho}^{k}\Lambda^{+}(x^{(\eta)}_{\rho}) \vert \lambda_{1},\lambda_{2}\rangle \nonumber\\
& &+\sum_{i}^{k}\sum^{k}_{i'\,(\neq i)}a(x_{i'}^{(\eta)},x_{i}^{(\eta)} )\,\Lambda^{+}(g)\prod_{\rho\,(\neq i)}^{k}\Lambda^{+}(x^{(\eta)}_{\rho})
\vert \lambda_{1},\lambda_{2}\rangle \nonumber\\
& & + \sum^{k}_{i}\sum^{k}_{i'=i+1}c(x_{i}^{(\eta)},x_{i'}^{(\eta)} )\prod_{\rho}^{k}\Lambda^{+}(x^{(\eta)}_{\rho})\vert \lambda_{1},\lambda_{2}\rangle.
\end{eqnarray}
It is clear that the second and the fourth terms in (\ref{15})
are proportional to the Bethe ansatz state (\ref{su11ba}),
which is assumed to be the eigenstate of the system.
Therefore, the eigen-energy of the 2A2S Hamiltonian is given by
\begin{eqnarray}\label{17}
E^{(\eta)}_{k,\mu_{1},\mu_{2}}=\chi
\sum_{i=1}^{k}\left(
\beta(x^{(\eta)}_{i})+\sum^{k}_{i'=i+1}c(x_{i}^{(\eta)},x_{i'}^{(\eta)} )\right),
\end{eqnarray}
as long as the first and the third terms proportional to
$\Lambda^{+}(g)\prod_{\rho\,(\neq i)}^{k}\Lambda^{+}(x^{(\eta)}_{\rho}) \vert \lambda_{1},\lambda_{2}\rangle$
in (\ref{15}) for given $i$ are cancelled out, which leads to the corresponding
equations in determining the $k$ variables $\{x_{1}^{(\eta)},\, x_{2}^{(\eta)},\cdots,\,x_{k}^{(\eta)}\}$:
\begin{equation}\label{BAE}
\alpha(x^{(\eta)}_{i})+\sum^{k}_{i'\,(\neq i)}a(x_{i'}^{(\eta)},x_{i}^{(\eta)} )=0
~~{\rm for}~~i=1,2,\cdots k.
\end{equation}

Using $\alpha(x^{(\eta)}_{i})$, $\beta(x^{(\eta)}_{i})$,
$a(x_{i'}^{(\eta)},x_{i}^{(\eta)} )$ and $c(x_{i}^{(\eta)},x_{i'}^{(\eta)} )$
in (\ref{g1p}) and (\ref{g2p}),
the eigen-energy  can be  simplified to
\begin{eqnarray}\label{171}
E^{(\eta)}_{k,\mu_{1},\mu_{2}}=2\,\chi\left(
\sum_{i=1}^{k}{\lambda_{2}\,g\, x^{(\eta)}_{i}-\lambda_{1}\over{g-x^{(\eta)}_{i}}}
-g\sum_{i=1}^{k}\sum^{k}_{i'=i+1}{1+x_{i}^{(\eta)}x_{i'}^{(\eta)}\over{(g-x^{(\eta)}_{i})(g-x^{(\eta)}_{i'})}}
\right),
\end{eqnarray}
where the $k$ variables $\{x_{i}^{(\eta)}\}$ should satisfy
\begin{equation}\label{BAE2}
{\lambda_{1}\over{x^{(\eta)}_{i}}}-\lambda_{2}\,x^{(\eta)}_{i}+
\sum^{k}_{i'\,(\neq i)}{1+x_{i}^{(\eta)}x_{i'}^{(\eta)}\over{x^{(\eta)}_{i}-x^{(\eta)}_{i'}}}=0
~~{\rm for}~~i=1,2,\cdots k,
\end{equation}
under the condition that $g\neq x^{(\eta)}_{i}~\forall~ \eta,\,i$.
It is obvious that $g= x^{(\eta)}_{i}$ for any $\eta$ and $i$ is
the singular point of (\ref{171}) and should be avoided.
It can be verified that root components $\{ x^{(\eta)}_{i}\}$ of (\ref{BAE2}), which are
always real and unequal one another, lie in the two open intervals $(-\infty,0)\bigcup(0,+\infty)$.

\vskip .3cm
By using (\ref{BAE2}), (\ref{171}) can be  expressed as
\begin{eqnarray}\label{1710}
E^{(\eta)}_{k,\mu_{1},\mu_{2}}=2\,\chi\left(
\sum_{i=1}^{k}{\lambda_{1}\over{x^{(\eta)}_{i}}}
+{1\over{2}}\sum_{i=1}^{k}\sum^{k}_{i'(\neq~i)}
{1\over{x^{(\eta)}_{i}-x^{(\eta)}_{i'}}}
{(1+x_{i}^{(\eta)}x_{i'}^{(\eta)})
(2g-x^{(\eta)}_{i}-x^{(\eta)}_{i'}) \over{
(g-x^{(\eta)}_{i})(g-x^{(\eta)}_{i'})}}
\right).
\end{eqnarray}
It is obvious that the first part $1/(x^{(\eta)}_{i}-x^{(\eta)}_{i'})$
within the double sum over $i$ and $i'$  in (\ref{1710})  is antisymmetric,
while the last part
is symmetric with respect to the permutation $i\rightleftharpoons i'$,
which ensures that the second term in (\ref{1710}) is zero.
Hence, the eigen-energies (\ref{171}) are
independent of $g$ as long as $g\neq x^{(\eta)}_{i}~\forall~ \eta,\,i$,
and can be further simplified to the form
\begin{equation}\label{1711}
E^{(\eta)}_{k,\mu_{1},\mu_{2}}=2\,\chi
\sum_{i=1}^{k}{\lambda_{1}\over{x^{(\eta)}_{i}}}=2\,\chi
\sum_{i=1}^{k}{\lambda_{2}{x^{(\eta)}_{i}}}
.
\end{equation}
%\red{Since zero is the singular point of the Bethe ansatz equations (\ref{BAE2}), namely, $x^{(\eta)}_{i}\neq0~\forall~ \eta,\,i$,  $g=0$ can be set in the very beginning shown in (\ref{g1})-(\ref{g2p}) to simplify the numeric.} Moreover,
It is now clear that $\eta$ labels the $\eta$-th solution of (\ref{BAE2}).
In addition, though the eigenstates provided in (\ref{su11ba}) are not normalized, they are always orthogonal with
\begin{equation}
 \langle \eta^{\prime}, k^{\prime};\lambda^{\prime}_{1},\lambda^{\prime}_{2}
\vert \eta, k;\lambda_{1},\lambda_{2}\rangle=({\cal N}(\eta,k;\lambda_{1},\lambda_{2}))^{-2}
\delta_{\eta \eta^{\prime}}\delta_{k k^{\prime}}
 \delta_{\lambda_{1} \lambda_{1}^{\prime}} \delta_{\lambda_{2} \lambda_{2}^{\prime}},
\end{equation}
where ${\cal N}(\eta,k;\lambda_{1},\lambda_{2})$ is the corresponding normalization constant.

\vskip .3cm
It is obvious that the Hamiltonian (\ref{1}) is invariant under the
permutations of the two spin operators
with $S^{\nu}_{1}\leftrightarrow S^{\nu}_{2}$
for $\nu=0,\,+,\,-$, which corresponds to the permutation of $a$-bosons with $b$-bosons.
The ${\rm SU}(1,1)$ operators $\Lambda^{+}(x_{i})$ used in (\ref{op}) are invariant
under the permutation of $a$-bosons with $b$-bosons.
Therefore, the SU(1,1) lowest weight states $\vert\lambda_{1},\lambda_{2}\rangle$
should be invariant under the permutation of $a$-bosons with $b$-bosons.
As clearly shown in Table \ref{t1}, the first two and the last two
two-spin intrinsic states
$\left\vert \begin{array}{l}
~~S_{1}^{\rm in}\\\
 M_{1}^{\rm in}
 \end{array} \begin{array}{l}
 S_{2}^{\rm in}\\
  M_{2}^{\rm in}
  \end{array}\right\rangle$
and $\left\vert \begin{array}{l}
~~S_{2}^{\rm in}\\\
 M_{2}^{\rm in}
 \end{array} \begin{array}{l}
 S_{1}^{\rm in}\\
  M_{1}^{\rm in}
  \end{array}\right\rangle$
are indeed the same SU(1,1) lowest weight state $\vert\lambda_{1},\lambda_{2}\rangle$
when both $\mu_{1}$ and $\mu_{2}\neq0$. The two-spin intrinsic state is unique with
$\left\vert \begin{array}{l}
~~S_{1}^{\rm in}\\\
 M_{1}^{\rm in}
 \end{array} \begin{array}{l}
 S_{2}^{\rm in}\\
  M_{2}^{\rm in}
  \end{array}\right\rangle=\left\vert \begin{array}{l}
~0\\\
 0
 \end{array} \begin{array}{l}
 0\\
  0
  \end{array}\right\rangle$
  only when $\mu_{1}=\mu_{2}=0$.
It is also obvious that the two-spin intrinsic states
with the same $M^{\rm in}=M_{1}^{\rm in}+M_{2}^{\rm in}$ value
are the same SU(1,1) lowest weight state $\vert\lambda_{1},\lambda_{2}\rangle$.

\vskip .3cm
Once the Bethe ansatz equations (\ref{BAE2}) are solved, the eigenstates (\ref{su11ba}), up to the two-spin permutation and a normalization constant,
can be expressed in terms of uncoupled two-spin states as
\begin{equation}\label{estate}
\vert \eta, k;\lambda_{1},\lambda_{2} \rangle=
\sum_{\rho=0}^{k}S_{\rho}^{(k,\eta)}\sqrt{{(k+\mu_{1}-\rho)!(\mu_{2}+\rho)!(k-\rho)!\rho!
%\over{\mu_{1}!\mu_{2}!}
}}
\left(
\begin{array}{l}
\left\vert
\begin{array}{l}
~~~{1\over{2}}(k+\mu_{1}+\mu_{2})~~~~~~~{k\over{2}}\\
{1\over{2}}(k+\mu_{1}-\mu_{2})-\rho~~{k\over{2}}-\rho
\end{array}
\right\rangle\\
\\
\left\vert
\begin{array}{l}
~~~{1\over{2}}(k+\mu_{1})~~~~~~~~{1\over{2}}(k+\mu_{2})\\
{1\over{2}}(k+\mu_{1})-\rho~~~{1\over{2}}(k-\mu_{2})-\rho
\end{array}
\right\rangle
\\
\end{array}
\right)
,
\end{equation}
where
\begin{equation}\label{Sq}
S^{(k,\eta)}_{0}=1,~~S^{(k,\eta)}_{q\geq1}=\sum_{1\leq\mu_{1}\neq\cdots\neq\mu_{q}\leq k}x^{(\eta)}_{\mu_{1}}
\cdots x^{(\eta)}_{\mu_{q}}
\end{equation}
are symmetric functions of $\{x^{(\eta)}_{1},\cdots,\,x^{(\eta)}_{k}\}$.

\vskip .3cm

Moreover, it can be verified directly that  (\ref{BAE2}) and the corresponding eigen-energy (\ref{171})
are invariant under the simultaneous interchanges
$\lambda_{1}\leftrightarrow\lambda_{2}$ and $x^{(\eta)}_{i}\leftrightarrow 1/x^{(\eta)}_{i}$, which
corresponds to the permutation of the two copies of the SU(1,1) generators,
$\Lambda^{\mu}_{1}\leftrightarrow\Lambda^{\mu}_{2}$ for $\mu=+,-,0$.
It is obvious that the 2A2S Hamiltonian (\ref{3}) is also invariant
under the  permutation $\Lambda^{\mu}_{1}\leftrightarrow\Lambda^{\mu}_{2}$.
Therefore, when $\lambda_{1}\neq\lambda_{2}$, the eigenvalue of the
2A2S Hamiltonian with $\{x_{i}^{(\eta)}\}$ built on the
SU(1,1) lowest weight state $\vert\lambda_{1},\lambda_{2}\rangle$
and that with $\{(x_{i}^{(\eta)})^{-1}\}$ built on the
$\vert\lambda_{2},\lambda_{1}\rangle$ are the same.
Thus, when $\lambda_{1}=\lambda_{2}$, if $\{x_{i}^{(\eta)}\}$
is a solution, $\{(x_{i}^{(\eta)})^{-1}\}$ gives the same
solution. Namely, for fixed $i$, if $x_{i}^{(\eta)}$ is one of the root
component, $x^{(\eta)}_{i'}=(x_{i}^{(\eta)})^{-1}$ is also a root
component of the same root. We can also verify that the roots of (\ref{BAE2}) have
the mirror symmetry. Namely, if $\{x_{i}^{(\eta)}\}$ is
a solution, $\{-x_{i}^{(\eta)}\}$ is also a solution. Thus, if the eigen-energy  is nonzero, the sign of the eigen-energy with $\{x_{i}^{(\eta)}\}$
is opposite to that with $\{-x_{i}^{(\eta)}\}$.

\vskip .3cm
According to the Heine-Stieltjes correspondence \cite{pan2019,3,4,5},
the second-order Fuchsian equation of the extended Heine-Stieltjes polynomials whose
zeros are the roots of (\ref{BAE2}) can be established. By using the identity
\begin{equation}
\sum_{i'(\neq i)}{x_{i'}^{(\eta)}\over{x_{i}^{(\eta)}-x_{i'}^{(\eta)}}}=
x_{i}^{(\eta)}\sum_{i'\neq i}{1\over{x_{i}^{(\eta)}-x_{i'}^{(\eta)}}}-k+1,
\end{equation}
it can be verified that the related extended Heine-Stieltjes polynomials $y^{{(\eta)}}_{k}(x)$
of degree $k$   should satisfy
\begin{equation}\label{fe}
(x+x^3)\,{d^{2}y_{k}^{{(\eta)}}(x)\over{dx^{2}}}+\left({2\lambda_{1}}-2({\lambda_{2}+k-{1}})\,x^2
\right){dy^{{(\eta)}}_{k}(x)\over{dx}}+{V^{{(\eta)}}(x)}\,y^{{(\eta)}}_{k}(x)=0,
\end{equation}
where the Van Vleck polynomial $V^{{(\eta)}}(x)$ is simply a binomial to be determined by (\ref{fe}).
Write $y^{(\eta)}_{k}(x)=\sum_{n=0}^{k}f^{(\eta)}_{n}x^{n}$ and
$V^{{(\eta)}}(x)=v^{(\eta)}_{0}+v^{(\eta)}_{1} x$,
where $\eta$ labels the $\eta$-th polynomial. It can be verified directly
that  the expansion coefficients $f^{(\eta)}_n$ ($n=0,\cdots, k$) should satisfy
the following three-term recurrence relations:
\begin{equation}\label{re}
(n+1)(2\lambda_{1}+n)f^{(\eta)}_{n+1}+v^{(\eta)}_{0}\,f^{(\eta)}_{n}+\left(
(n-1)(n-2\lambda_{2}-2k)+v^{(\eta)}_{1}\right)f^{(\eta)}_{n-1}=0
\end{equation}
with
\begin{equation}\label{re1}
v^{(\eta)}_{1}=k ( 2 \lambda_{2}+  k-1),
\end{equation}
which is independent of $\eta$.
Instead of solving the three-term recurrence relations (\ref{re})
for the expansion coefficients $f^{(\eta)} _n$ ($n=0,\cdots, k$) and $v^{(\eta)}_{0}$, we can construct the corresponding $(k+1)\times(k+1)$ bidiagonal matrix $A$ with entries
 \begin{equation}\label{re2}
A_{n,n'}=\left((n-1)(n-2\lambda_{2}-2k)+k ( 2 \lambda_{2}+  k-1)\right)\delta_{n',n-1}+
(n+1)(2\lambda_{1}+n)\delta_{n',n+1}.
\end{equation}
Let ${\bf F}^{(\eta)}=(f^{(\eta)}_{0},\,f^{(\eta)}_{1},\,\cdots,\,f^{(\eta)}_{k})^{\bf T}$,
where ${\bf T}$ stands for the transpose operation.
The $k+1$ dimensional vector ${\bf F}^{(\eta)}$ is the $\eta$-th eigenvector
of the bidiagonal matrix $A$ with
 \begin{equation}\label{eign}
A{\bf F}^{(\eta)}=(-v^{(\eta)}_{0}){\bf F}^{(\eta)},
\end{equation}
where $-v^{(\eta)}_{0}$ is the corresponding eigenvalue, {which
clearly shows that there are $k+1$ sets of solutions of (\ref{BAE2})
for the eigen-energies (\ref{1711}) and the corresponding eigenstates
(\ref{estate}) with $\eta=1,\,2,\,\cdots,\,k+1$.} It is also obvious that not only
the construction of the matrix $A$,
but also its diagonalization is easier with smaller size of
the matrix and thus more efficient
than the direct diagonalization of the 2A2S Hamiltonian (\ref{1})
in the original uncoupled two-spin  basis, especially when $k$ is large.
Once the $\eta$-th set of the expansion coefficients
$f^{(\eta)} _n$ ($n=0,\cdots, k$) are known from
the eigen-equation (\ref{eign}), the $k$ zeros
$\{x_{i}^{(\eta)}\}$ ($i=0,\cdots, k$) of the polynomial
{$y^{(\eta)}_{k}(x)/{f}^{(\eta)}_{k}=\sum_{n=0}^{k}\bar{f}^{(\eta)}_{n}x^{n}$,
where, up to an overall factor, $\bar{f}^{(\eta)}_{n}= {f}^{(\eta)}_{n}/{f}^{(\eta)}_{k}$
($n=1,\cdots,k$), can easily be calculated due to the fact that
$y^{(\eta)}_{k}(x)$ is one-variable polynomial and ${f}^{(\eta)}_{k}$ is always nonzero.
In addition, $y^{(\eta)}_{k}(x)/{f}^{(\eta)}_{k}$  can also be expressed in terms of the zeros
$\{x^{(\eta)}_{j}\}$ ($j=1,\cdots,k$)  as
\begin{equation}\label{20}
y^{(\eta)}_{k}(x)/{f}^{(\eta)}_{k}=\prod_{j=1}^{k}(x-x^{(\eta)}_{j})=\sum_{q=0}^{k}(-1)^{q} {S}_{q}^{(k,\eta)}x^{k-q},
\end{equation}
where  ${S}_{q}^{(k,\eta)}$ is the symmetric function defined in (\ref{Sq}).
Comparing (\ref{20}) with
$y^{(\eta)}_{k}(x)/{f}^{(\eta)}_{k}=\sum_{n=0}^{k}\bar{f}^{(\eta)}_{n}x^{n}$, we get
\begin{equation}\label{21}
{S}^{(k,\eta)}_{q}=(-1)^{q} \bar{f}^{(\eta)}_{k-q}.
\end{equation}
} 
which can be used  to avoid
unnecessary computation of $S^{(k,\eta)}_{q}$
from $\{x^{(\eta)}_{1},\cdots,~x^{(\eta)}_{k}\}$
needed in the eigenstates (\ref{estate}).
Furthermore, using the three-term recurrence relations (\ref{re}), we have
\begin{equation}\label{211}
v_{0}^{(\eta)}=-2\lambda_{1}{\bar{f}^{(\eta)}_{1} /{\bar{f}^{(\eta)}_{0} }}=
2\lambda_{1}S^{(k,\eta)}_{k-1}/S^{(k,\eta)}_{k}=E^{(\eta)}_{k,\mu_{1},\mu_{2}}/\chi,
\end{equation}
where the relation (\ref{21}) is used for the second equality, which shows that
the constant term $v_{0}^{(\eta)}$ in the Van Vleck polynomial
equals exactly to the corresponding eigen-energy $E^{(\eta)}_{k,\mu_{1},\mu_{2}}/\chi$
of the 2A2S Hamiltonian.

\vskip .3cm
{Finally, we prove that  the solutions of the
Bethe ansatz equations (\ref{BAE2}) are complete.
When the two spins are unequal with $S_1>S_{2}$,
there are three sets of uncoupled two-spin states
used in the eigenstates (\ref{estate}):
Case 1 with $S_{1}={1\over{2}}(k+\mu_{1}+\mu_{2})$ and $S_{2}={1\over{2}}k$
corresponding to the upper one in (\ref{estate});
Case 2 with $S_{1}={1\over{2}}(k+\mu_{1})$ and $S_{2}={1\over{2}}(k+\mu_{2})$
corresponding to the lower one in (\ref{estate});
and Case 3 with $S_{1}={1\over{2}}(k+\mu_{2})$ and  $S_{2}={1\over{2}}(k+\mu_{1})$ 
corresponding to the two-spin permutation $S_{1}\rightleftharpoons S_{2}$
of the lower one in (\ref{estate}).
\vskip .3cm
For Case 1, $S_{1}={1\over{2}}(k+\mu_{1}+\mu_{2})>S_{2}={1\over{2}}k$
with
$\mu_{1}=2S_{1}-2S_{2}-\mu_{2}\geq0$, where
$\mu_{2}$ can be taken as $0,1,2\cdots,2S_{1}-2S_{2}$. 
The number of solutions provided by (\ref{BAE2}) for a
fixed $\mu_{2}$ is $k+1=2S_{2}+1$.
Thus, the total number of solutions for $S_{1}={1\over{2}}(k+\mu_{1}+\mu_{2})$ and $S_{2}={1\over{2}}k$ case shown by (\ref{estate}) with 
the upper uncoupled two-spin states
is $(2S_{2}+1)(2S_{1}-2S_{2}+1)$, which includes $\mu_{2}=0$ case.
\vskip .3cm
For Case 2,  $S_{1}={1\over{2}}(k+\mu_{1})>S_{2}={1\over{2}}(k+\mu_{2})$
with $\mu_{1}=\mu_{2}+2S_{1}-2S_{2}$ and $\mu_{2}>0$
because $\mu_{2}=0$ case is already
considered in Case 1,
the number of solutions provided by (\ref{BAE2}) is $k+1=2S_{2}-\mu_{2}+1$
for a fixed $\mu_{2}$,
where $\mu_{2}$ can be taken as $1,2,\cdots,2S_{2}$.
Hence, the total number of solutions for this
case shown by (\ref{estate}) with the lower uncoupled two-spin states   
is $\sum_{\mu_{2}=1}^{2S_{2}}(2S_{2}-\mu_{2}+1)=
(2S_{2}+1)S_{2}$, which excludes the $\mu_{2}=0$ case.
\vskip .3cm
For Case 3, $S_{1}={1\over{2}}(k+\mu_{2})>S_{2}={1\over{2}}(k+\mu_{1})$
corresponding to the two-spin permutation $S_{1}\rightleftharpoons S_{2}$
of the lower one in (\ref{estate}),
where $\mu_{1}=\mu_{2}-2S_{1}+2S_{2}\geq0$ and $2S_{1}\geq\mu_{2}\geq 2S_{1}-2S_{2}$,
the number of solutions provided by (\ref{BAE2}) is $k+1=2S_{1}-\mu_{2}+1$ for a fixed
$\mu_{2}$. Since $\mu_{2}=2S_{1}-2S_{2}$ with $\mu_{1}=0$ case is already considered in Case 2,
the total number of solutions for this
case is $\sum_{\mu_{2}=2S_{1}-2S_{2}+1}^{2S_{1}}(2S_{1}-\mu_{2}+1)=
(2S_{2}+1)S_{2}$, which excludes the $\mu_{2}=2S_{1}-2S_{2}$ case.
\vskip .3cm
It is now obvious that the total number of the solutions
provided by the three cases equals exactly to
$(2S_{1}+1)(2S_{2}+1)$, which is
the dimension of the uncoupled two-spin states
for $S_{1}>S_{2}$.
This conclusion also applies to $S_{1}<S_{2}$ case by 
the  permutation $S_{1}\rightleftharpoons S_{2}$.

For the two equal spin case with $S_{1}=S_{2}=k/2$,
which is exemplified in the next section,
the upper and lower uncoupled two-spin states
shown in (\ref{estate})
are the same.
According to (\ref{estate}), the two-spin symmetric or anti-symmetric eigenstates
 of the 2A2S Hamiltonian  in this case can be written uniformly as
\begin{eqnarray}\label{estate1}
&&\vert \eta, k-\mu;(\mu+1)/2,(\mu+1)/2 \rangle_{\rm S,A}\nonumber\\
&&\qquad\qquad =\sum_{\rho=0}^{k-\mu}S_{\rho}^{(k-\mu,\eta)}\sqrt{(k-\rho)!(\mu+\rho)!(k-\mu-\rho)!\rho!}
\begin{array}{l}
\left\vert
\begin{array}{l}
~~~{k\over{2}}~~~~~~~~~{k\over{2}}\\
{k\over{2}}-\rho~~{k\over{2}}-\mu-\rho
\end{array}
\right\rangle_{\rm S,A}\\
\end{array}
\end{eqnarray}
for $\mu=0,\,1,\,\cdots,\,k$,
where
\begin{eqnarray}\label{estate11}
&\begin{array}{l}
\left\vert
\begin{array}{l}
~~~{k\over{2}}~~~~~~~~~{k\over{2}}\\
{k\over{2}}-\rho~~{k\over{2}}-\mu-\rho
\end{array}
\right\rangle_{\rm S}=\sqrt{1\over{2(1+\delta_{\mu0})}}\left(
\left\vert
\begin{array}{l}
~~~{k\over{2}}~~~~~~~~~{k\over{2}}\\
{k\over{2}}-\rho~~{k\over{2}}-\mu-\rho
\end{array}
\right\rangle
+
\left\vert
\begin{array}{l}
~~~~~{k\over{2}}~~~~~~~~~~{k\over{2}}\\
{k\over{2}}-\mu-\rho~~{k\over{2}}-\rho
\end{array}
\right\rangle
\right),
\end{array}
\end{eqnarray}
%and
\begin{eqnarray}\label{estate111}
&\begin{array}{l}
\left\vert
\begin{array}{l}
~~~{k\over{2}}~~~~~~~~~{k\over{2}}\\
{k\over{2}}-\rho~~{k\over{2}}-\mu-\rho
\end{array}
\right\rangle_{\rm A}=\sqrt{1\over{2}}\left(
\left\vert
\begin{array}{l}
~~~{k\over{2}}~~~~~~~~~{k\over{2}}\\
{k\over{2}}-\rho~~{k\over{2}}-\mu-\rho
\end{array}
\right\rangle
-
\left\vert
\begin{array}{l}
~~~~~{k\over{2}}~~~~~~~~~~{k\over{2}}\\
{k\over{2}}-\mu-\rho~~{k\over{2}}-\rho
\end{array}
\right\rangle
\right)
\end{array}
\end{eqnarray}
are the symmetric and anti-symmetric  two-spin  states, respectively.
The number of solutions of (\ref{BAE2}) with the replacement: $k\rightarrow k-\mu$ 
with a fixed $\mu$  for (\ref{estate1}) is $k-\mu+1$. Due to the two-fold
degeneracy of the $\mu\neq 0$  states, the total number of solutions
provided by  (\ref{estate1}) is $2\sum_{\mu=1}^{k}(k-\mu+1)+k+1=(k+1)^2$.}

\section{Some numerical examples of the solution}

To demonstrate the method and results presented above, in this section, we
consider the two equal spin case with $S_{1}=S_{2}=k/2$ as studied in \cite{2222}.
The two-spin symmetric or anti-symmetric eigenstates
of the 2A2S Hamiltonian  are given by (\ref{estate1}).
It is obvious that the $\mu=0$ eigenstates  of the 2A2S Hamiltonian are always symmetric,
while both symmetric and anti-symmetric states are possible for $\mu\neq0$.
Moreover, the corresponding eigen-energies $E^{(\eta)}_{k-\mu,\mu,\mu}/\chi$ ($\eta=1,2,\cdots,k-\mu+1$) of both two-spin symmetric and anti-symmetric cases
are symmetric with respect to the the sign change.
Namely, if the $k-\mu+1$ level energies are arranged as
$E^{(1)}_{k-\mu,\mu,\mu}/\chi<E^{(2)}_{k-\mu,\mu,\mu}/\chi<\cdots
<E^{(k-\mu+1)}_{k-\mu,\mu,\mu}/\chi$, then
\begin{eqnarray}
E^{(k-\mu+2-q)}_{k-\mu,\mu,\mu}/\chi=-E^{(q)}_{k-\mu,\mu,\mu}/\chi
\end{eqnarray}
for
\begin{eqnarray}
q=\left\{
\begin{array}{l}
1,2,\cdots,(k-\mu+1)/2~~\rm{when}~~k-\mu+1~\rm{is~even},\\
1,2,\cdots,(k-\mu+2)/2~~\rm{when}~~k-\mu+1~\rm{is~odd},
\end{array}\right.
\end{eqnarray}
which shows  that the middle level energy $E^{((k-\mu+2)/2)}_{k-\mu,\mu,\mu}=0$
when $k-\mu+1$ is odd. As the consequence, for given $k$,
the degeneracy of the zero eigen-energy level is $k+1$.

\begin{table*}[h]
\caption{The Heine-Stieltjes Polynomials $y^{(\zeta)}_{k-\mu}(x)$,
 $v^{(\eta)}_{0}$ of the corresponding Van Vleck Polynomial $V^{(\zeta)}(x)$,
 and the corresponding eigen-energy $E^{(\eta)}_{k-\mu,\mu,\mu}/\chi$  of the 2A2S
 Hamiltonian (1) for $S_{1}=S_{2}=k/2$ and $k\leq4$, where the ordering of $\eta$ is arranged according to
 the value of the eigen-energy of (1) for a given set of $\{ k,\mu\}$.}
\begin{tabular}{ccccc}
\hline\hline
$k,\mu$&$\eta$ &{$y^{(\eta)}_{k-\mu}(x)$} &$v^{(\eta)}_{0}$=$E^{(\eta)}_{k-\mu,\mu,\mu}/\chi$ \\
\hline
\hline
1,~0 &$1$ &$x+1$ &$-1$ \\
    &$2$ &$x-1$ &$~~1$ \\
~~~1 &$1$ &$1$ &$~~0$ \\

2,~0 &$1$ &$(x+0.4142) (x+2.4142)$ &$-2.8284$ \\
    &$2$ &$(x+1)(x-1)$ &$~~0$ \\
    &$3$ &$(x-2.4142) (x-0.4142)$ &~~$2.8284$ \\
~~~1 &$1$ &$x+1$ &$-2$ \\
     &$2$ &$x-1$ &$~~2$ \\
~~~2 &$1$ &$1$ &$~~0$ \\
3,~0 &$1$ &$(x+0.2285)(x+1)(x+4.3771)$ &$-5.6056$ \\
     &$2$ &$(x-1)(x+0.4678)(x+2.1378)$ &$-1.6056$ \\
     &$3$ &$(x-2.1378)(x-0.4678)(x+1)$ &$~~1.6056$ \\
     &$4$ &$(x-4.3771)(x-1)(x-0.2285)$ &$~~5.6056$ \\
~~~1 &$1$ &$(x+0.5176)(x+1.9319)$ &$-4.8989$ \\
     &$2$ &$(x-1)(x+1)$ &$~~~0$ \\
     &$3$ &$(x-1.9319)(x-0.5176)$ &~~$4.8989$ \\
  ~~2 &$1$ &$x+1$ &$-3$ \\
     &$2$ &$x-1$ &$~~3$ \\
  ~~3 &$1$ &$1$ &$~~0$ \\
4,~0 &$1$ &$(x+0.1429)(x+0.6151) (x+1.6259) (x+6.9970)$ &$-9.3808$ \\
     &$2$ &$(x-1) (x+0.2679)(x+1)(x+3.7321)$ &$-4$ \\
     &$3$ &$(x-1.9319) (x-0.5176) (x+0.5176) (x+1.9319)$ &$~~0$ \\
     &$4$ &$(x-3.7321)(x-1)(x-0.2679)(x+1) $ &$~~4$ \\
     &$5$ &$ (x-6.9970)(x-1.6259)(x-0.6151)(x-0.1429) $ &~~$9.3808$ \\
  ~~1 &$1$ &$(x+0.3285) (x+1) (x+3.0437)$ &$-8.7444$ \\
     &$2$ &$(x-1) (x+0.5482)(x+1.8241)$ &$-2.7446$ \\
     &$3$ &$(x-1.8241) (x-0.5482)(x+1)$ &~~$2.7446$ \\
     &$4$ &$(x-3.0437)(x-1)(x-0.3285)  $ &~~$8.7444$ \\
 ~~2 &$1$ &$(x+0.5774) (x+1.7321)$ &$-6.9282$ \\
     &$2$ &$(x-1) (x+1)$ &$0$ \\
     &$3$ &$ (x-1.7321)(x-0.5774)$ &~~$6.9282$ \\
 ~~3 &$1$ &$x+1$ &$-4$ \\
     &$2$ &$x-1$ &~~$4$ \\
 ~~4 &$1$ &$1$ &~~$0$ \\
    \hline\hline
\end{tabular}\label{t2}
\end{table*}

\begin{center}\begin{table}
\centering\textwidth1.038pt\tabcolsep0.032in\fontsize{8.pt}{8.pt}\selectfont
\caption{The same as Table \ref{t2}, but for $k=16$ and $\mu=0$.}
\begin{tabular}{ccccc}
\hline\hline
&$\eta$ &{$y^{(\eta)}_{16}(x)$} &$v^{(\eta)}_{0}$=$E^{(\eta)}_{16,0,0}/\chi$ \\
\hline
\hline
 &$1$ &$(x+0.0108) (x+0.0553) (x+0.1298) (x+0.2288)(x+0.3497)(x+0.4933)(x+0.6648) (x+0.8749)$ & \\
    && $(x+1.1430)(x+1.5042) (x+2.0272)(x+2.8598) (x+4.3702) (x+7.7057)(x+18.0792) (x+92.3648)$ &-132.862\\\\
    &$2$ &$(x-1) (x+0.0130) (x+0.0656) (x+0.1515) (x+0.2635)
     (x+0.3988) (x+0.5600) (x+0.7550)$ & \\
    && $(x+1) (x+1.3245)(x+1.7858) (x+2.5073) (x+3.7948) (x+6.5992) (x+15.2406) (x+76.8867)$ &-110.346\\
    \\
    &$3$ &$(x-1.3550)(x-0.7380) (x+0.0160)(x+0.0791)(x+0.1787)(x+0.3055)(x+0.4574) (x+0.6396)$ & \\
    && $(x+0.8645) (x+1.15671)(x+1.5636)(x+2.1861)(x+3.2735)(x+5.5949)(x+12.6385)(x+62.5073)$ &-89.3684\\\\
    &$4$ &$(x-1.7302)(x-1)(x-0.5780) (x+0.0203)(x+0.09718) (x+0.2129)(x+0.3563)(x+0.5276)$ & \\
    && $(x+0.7356) (x+1)(x+1.3595)(x+1.8955)(x+2.806)(x+4.6972)(x+10.2903)(x+49.2734)$ &-69.9638\\
    \\
&$5$ &$(x-2.1755)(x-1.2622)(x-0.7922)(x-0.4597)(x+0.0268)(x+0.1217)(x+0.2558)(x+0.4180)$ & \\
    && $(x+0.6121)(x+0.8531)(x+1.1721)(x+1.6338)(x+2.3924)(x+3.9092)(x+8.2178)(x+37.2658)$ &-52.189\\
  \\
&$6$ &$(x-2.7484)(x-1.5548)(x-1)(x-0.6432)(x-0.3639)(x+0.0375)(x+0.1552)(x+0.3095)$ & \\
    && $(x+0.4928)(x+0.7148)(x+1)(x+1.3989)(x+2.0293)(x+3.2309)(x+6.4441)(x+26.642)$ &-36.145\\
  \\
 &$7$ &$(x-3.5500)(x-1.9054)(x-1.2235)(x-0.8173)(x-0.5248)(x-0.2817)(x+0.0564)(x+0.2007)$ & \\
    && $(x+0.3763)(x+0.5837)(x+0.8414)(x+1.1885)(x+1.7132)(x+2.6578)(x+4.9833)(x+17.7332)$ &-22.032\\
  \\
 &$8$ &$(x-4.8049)(x-2.3557)(x-1.4807)(x-1)(x-0.6753)(x-0.4245)(x-0.2081)(x+0.0900)$ & \\
    && $(x+0.2609)(x+0.4584) (x+0.6947)(x+1)(x+1.4395)(x+2.1814)(x+3.8322)(x+11.1066)$ &-10.114\\
\\
&$9$ &$(x-7.0260)(x-2.9726)(x-1.7931)(x-1.2039)(x-0.8307)(x-0.5577)(x-0.3364)(x-0.1423) $ & \\
    && $(x+0.1423)(x+0.3364)(x+0.5577)(x+0.8307)(x+1.2039) (x+1.7931)(x+2.9726)(x+7.0260)$ &~~0\\
 \\
&$10$ &$(x-11.1066)(x-3.8322)(x-2.1814)(x-1.4395)(x-1)(x-0.6947)(x-0.4584)(x-0.2609)$ & \\
    && $(x-0.0900)(x+0.2081)(x+0.4245)(x+0.6753)(x+1)(x+1.4807)(x+2.3557)(x+4.8049) $ &~~10.114\\
 \\
&$11$ &$(x-17.7332)(x-4.9833)(x-2.6578)(x-1.7132)(x-1.1885)(x-0.8414)(x-0.5837)(x-0.3763)$ & \\
    && $(x-0.2007)(x-0.0564)(x+0.2817)(x+0.5248)(x+0.8173)(x+1.2235)(x+1.9054)(x+3.5500)$ &~~22.032\\
 \\
&$12$ &$(x-26.642)(x-6.4441)(x-3.2309)(x-2.0293)(x-1.3989)(x-1)(x-0.7148)(x-0.4928)$ & \\
    && $(x-0.3095)(x-0.1552)(x-0.0375)(x+0.3639)(x+0.6432)(x+1)(x+1.5548)(x+2.7484)$ &~~36.145\\
\\
&$13$ &$(x-37.2658)(x-8.2178)(x-3.9092)(x-2.3924)(x-1.6338)(x-1.1721)(x-0.8531)(x-0.6121)$ & \\
    && $(x-0.4180)(x-0.2558)(x-0.1217)(x-0.0268)(x+0.4597)(x+0.7922)(x+1.2622)(x+2.1755)$ &~~52.189\\
    \\
&$14$ &$(x-49.2734)(x-10.2903)(x-4.6972)(x-2.806)(x-1.8955)(x-1.3595)(x-1)(x-0.7356)$ & \\
    && $(x-0.5276)(x-0.3563)(x-0.2129)(x-0.09718)(x-0.0203)(x+0.5780)(x+1)(x+1.7302)$ &~~69.9638\\
    \\
&$15$ &$(x-62.5073)(x-12.6385)(x-5.5949)(x-3.2735)(x-2.1861)(x-1.5636)(x-1.15671)(x-0.8645)$ & \\
    && $(x-0.6396)(x-0.4574)(x-0.3055)(x-0.1787)(x-0.0791)(x-0.0160)(x+0.7380)(x+1.3550)$ &~~89.3684\\
    \\
&$16$ &$(x-76.8867)(x-15.2406)(x-6.5992)(x-3.7948)(x-2.5073)(x-1.7858)(x-1.3245)(x-1)$ & \\
    && $(x-0.3988)(x-0.7550)(x-0.5600)(x-0.2635)(x-0.1515)(x-0.0656)(x-0.0130)(x+1)$ &~~110.346\\
    \\
&$17$ &$(x-92.3648)(x-18.0792)(x-7.7057)(x-4.3702)(x-2.8598)(x-2.0272)(x-1.5042)(x-1.1430)$ & \\
    && $(x-0.8749)(x-0.6648)(x-0.4933)(x-0.3497)(x-0.2288)(x-0.1298)(x-0.0553)(x-0.0108)$ &~~132.862\\
 \hline\hline
\end{tabular}\label{t3}
\end{table}
\end{center}
\vskip -.6cm

\begin{figure}[H]
\begin{center}
\caption{The level patten of the 2A2S Hamiltonian for $S_{1}=S_{2}=k/2$
with $k=4$, where the number on the right of each level is the degeneracy
due to the two-spin permutation symmetry, and the dashed line indicates
the $E=0$ plane perpendicular to the level diagram as a mirror, with which the levels in each band labelled by $\mu$ are mirror symmetric. }
\vskip.1in
\includegraphics[scale=0.6]{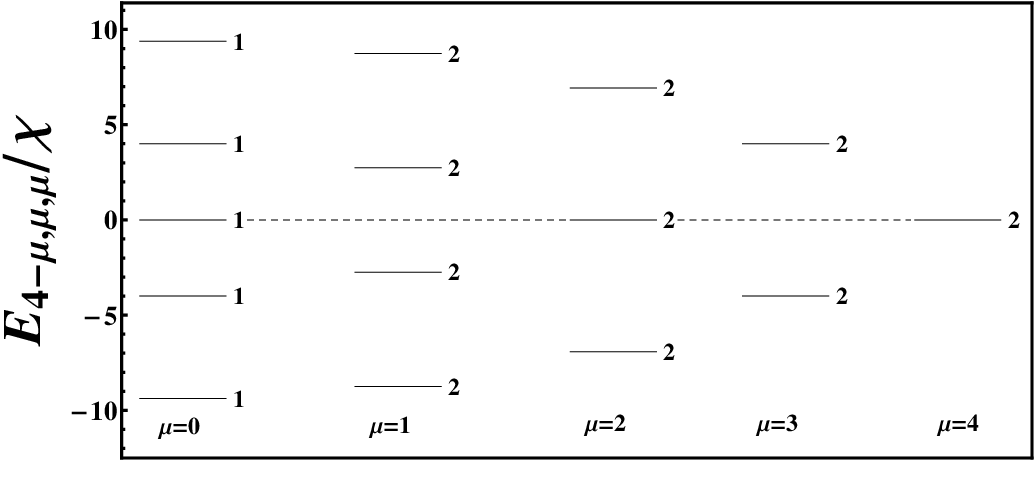}
\label{f1}
\end{center}
\end{figure}

\indent
The Heine-Stieltjes polynomials $y_{k-\mu}^{(\eta)}(x)$ and
the corresponding coefficient $v_{0}^{(\eta)}$ in the Van Vleck polynomials (\ref{17})
up to $k=4$ are shown in Table \ref{t2}, while the $k=16$ and $\mu=0$ case
is provided in Table \ref{t3}. It should be noted that the degeneracy of the
corresponding level energy of the 2A2S Hamiltonian
shown in the last column of Tables \ref{t2} and \ref{t3}
is $1$ for $\mu=0$ case and $2$ for $\mu\neq0$ case due to the
two-spin permutation symmetry. For any case, it can be verified that
any zero of $y_{{k-\mu}}^{(\eta)}(x)$ is always real and lies in one
of the intervals $(-\infty, 0)$ and $(0, \infty)$, which ensures
that eigenvalues (\ref{1711}) are always real.
Fig. \ref{f1} provides the level pattern of the 2A2S Hamiltonian
for $k=4$ with the level band labelled by $\mu$,
where the number on the right of each level is the degeneracy
due to the two-spin permutation symmetry, which clearly shows  that
the mirror symmetry of the $k-\mu+1$ levels
in each band with respect to the $E=0$ plane perpendicular to the level diagram
and that the total number of levels for given $k$
equals exactly to $(k+1)^2$, which is the total dimension of the two-spin basis.
In addition, when quantum numbers of the two spins of the system are small,
the 2A2S Hamiltonian can easily be  diagonalized
within the two-spin basis, with which
one can check that the eigen-energies shown in Tables \ref{t2} and \ref{t3}
are exactly the same as those obtained from the direct diagonalization, which validates the Bethe ansatz solution presented in Sec. II.

As an application of the solution,
the entanglement measure of all possible 2A2S pure states (\ref{estate1})
with $k=40$ is calculated, which, for given $\mu$ and $\eta$,
is quantified by the von Neumann entropy
\begin{equation}
{\rm Ent}(\mu,\eta)=-{\rm Tr}(\rho^{(\mu,\eta)}_{1}\,{\rm Log}_{N}\rho^{(\mu,\eta)}_{1})
=-{\rm Tr}(\rho^{(\mu,\eta)}_{2}\,{\rm Log}_{N}\rho^{(\mu,\eta)}_{2}),
\end{equation}
where $\rho^{(\mu,\eta)}_{1}$ or $\rho^{(\mu,\eta)}_{2}$ is the reduced
density matrix of the two-spin symmetric (S) or anti-symmetric (A) eigenstate (\ref{estate1})
obtained by taking the partial trace over the subsystem of spin $2$ or $1$, and
the logarithm to the base $N=2(k-\mu+1)$ for both the two-spin symmetric (S) and  anti-symmetric (A)
eigenstates with $\mu\neq0$ or $N=k+1$ for the two-spin symmetric (S) eigenstates with $\mu=0$, which is the total number of modes involved,
is used to ensure that the maximum measure is normalized to $1$.

\begin{figure}[H]
\begin{center}
\caption{(Color online)
The entanglement measure ${\rm Ent}(\mu,\eta)$ of all excited states
of the 2A2S system with $k=40$, where the green and red colors are
used to distinguish the measures of the excited states from the adjacent bands. }
\includegraphics[scale=0.5]{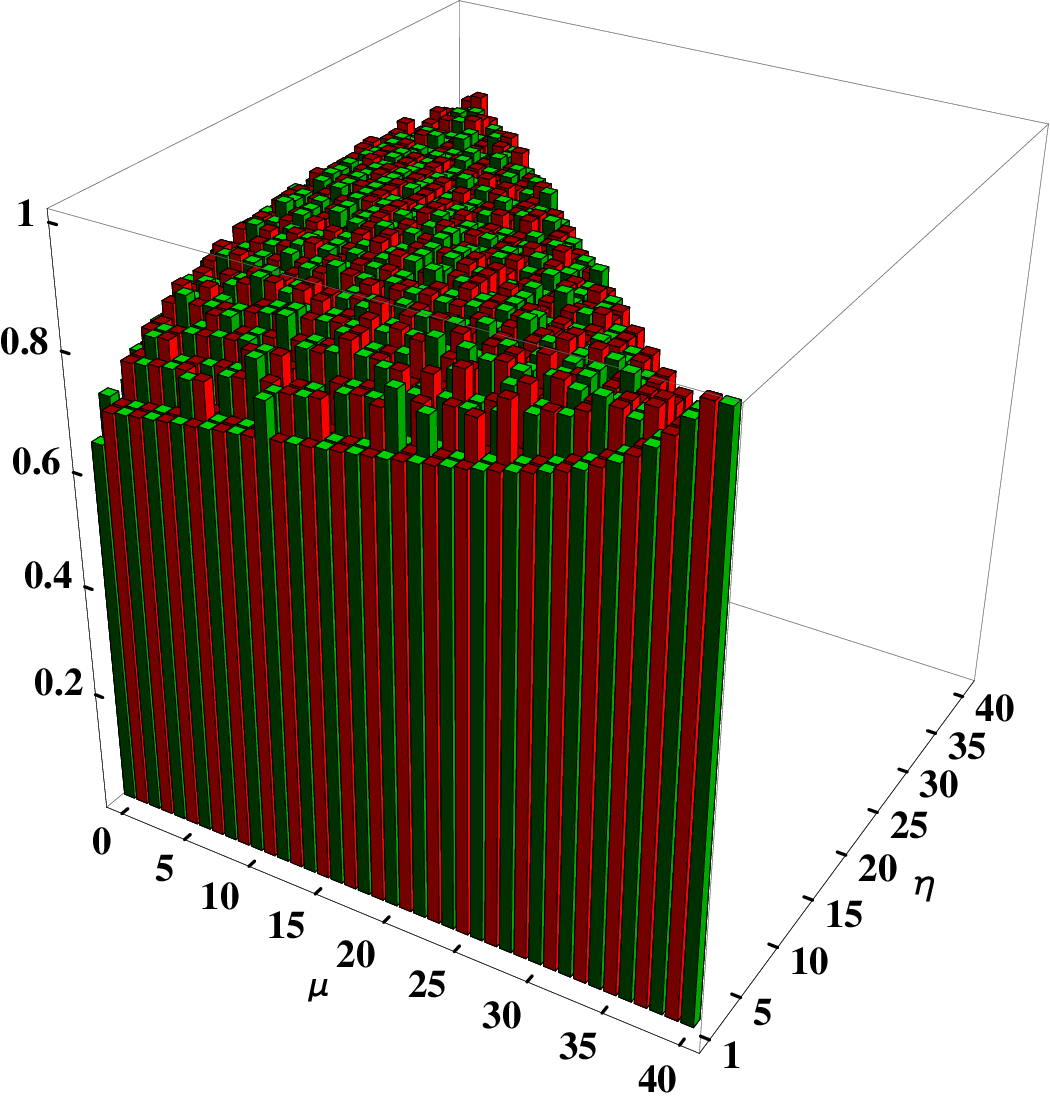}
\label{f2}
\end{center}
\end{figure}

Fig. \ref{f2} shows entanglement measure of all excited states
of the system  with $k=40$. It is observed that the excited states
are all well entangled with measure ${\rm Ent}(\mu,\eta)\geq 0.6476$
in the $k=40$ case. The entanglement measure for both the two eigenstates in the $(k-1)$-th band and
the single eigenstate in the $k$-th band reaches to the maximum value, i.e. ${\rm Ent}(k-1,\eta)=1$ for $\eta=1,\,2$ and ${\rm Ent}(k,1)=1$, respectively,  and the
excited states in other bands with $\mu\leq k-2$ are also
well entangled with $0.9671\geq{\rm Ent}(\mu,\eta)\geq 0.6476$.
Moreover, the entanglement measure of the band-head states
gradually increases with the increasing of its excitation energy or $\mu$
from ${\rm Ent}(0,1)=0.6476$ to ${\rm Ent}(40,1)=1$
in the $k=40$ case. Additionally, except the last two bands with $\mu=k-1$ or $k$,
the entanglement measure of the excited states in other bands
varies almost randomly with the value ${\rm Ent}(\mu,\eta)\geq{\rm Ent}(\mu,1)$
for $1\leq\eta\leq k-\mu+1$  as shown in Fig. \ref{f2}.
Therefore, the eigenstates of the 2A2S system are always well entangled.

\section{SUMMARY}

In this work, a systematic procedure for calculating exact solution of
the 2A2S Hamiltonian is presented by using the Bethe ansatz method.
It is shown that the 2A2S Hamiltonian can be expressed in terms of
the generators of two copies of SU(1,1) algebra after the Jordan-Schwinger
boson realization of the two related SU(2) algebras. Thus,  eigenstates of
the 2A2S Hamiltonian can be expressed as SU(1,1) type Bethe ansatz states, by which
eigen-energies and the corresponding eigenstates are derived.
To avoid solving a set of non-linear Bethe ansatz equations involved,
the related extended Heine-Stieltjes polynomials are constructed,
whose zeros are just components of a root of the Bethe ansatz equations.
Symmetry properties of excited levels of the 2A2S system
and those of zeros of the related extended Heine-Stieltjes polynomials
are also discussed. As an example, the two equal spin case is analysed in detail.
It is shown that the $(2S+1)^2$ dimensional energy matrix
in the  uncoupled two-spin basis, where $S=S_{1}=S_{2}$
is the quantum number of the two spins, can be decomposed into $2S-\mu+1$
dimensional bi-diagonal sub-matrices for $\mu=0,1,\cdots, 2S$
with two-fold degeneracy for $\mu\neq0$,
which clearly demonstrates the advantages of the Bethe ansatz
method over the direct diagonalization in the original uncoupled two-spin basis.
Furthermore, in the two equal spin case, it is shown that the levels in each band
labelled by $\mu$ are symmetric with respect to
the zero energy plane perpendicular to the level diagram
and that the excited states are always well entangled.

\vskip .3cm
{In comparison to the direct diagonalization of
the 2A2S Hamiltonian in the uncoupled two-spin basis with $S_{1}=S_{2}=k/2$,
the size of the energy matrix increases with $k$ quadratically,
while the size of the energy sub-matrices constructed by the Bethe ansatz
method shown in this paper increases with $k$ linearly, which makes
the diagonalization more efficient and doable for large spin cases.
For example, when the two spins are large with $k\sim 10^{10}$,
the diagonalization of the bidiagonal matrices 
with  $10^{10}$ in size can be carried out on a current day computer,
while the direct diagonalization of the energy matrix
in the original two-spin basis with  $10^{20}$ in size
becomes a formidable task. Therefore, the results shown in this paper should 
be useful for large spin cases in Bose-Einstein condensates \cite{bec}.

\vskip .3cm
Further analysis of the model using the Bethe ansatz solution, such as computing the overlaps of the eigenstates with the two-spin squeezed state as suggested in \cite{2222} and other physical quantities of the system, especially in the thermodynamic limit by using a similar procedure in  \cite{cr}, is beyond the scope of this paper and will be part of our future work to be presented elsewhere.}

\bigskip

\begin{acknowledgments}
{This work was supported by  National Natural Science Foundation of China (Grants No. 11675071 and No. 11775177), Australian Research Council Discovery Project DP190101529,
 U. S. National Science Foundation (OIA-1738287 and  PHY-1913728), and  LSU-LNNU joint research program with modest but important collaboration-maintaining support from the Southeastern Universities Research Association.}

\end{acknowledgments}

\end{document}